\documentclass[aip,jcp,groupedaddress,superscriptaddress,twocolumn]{revtex4}

\usepackage{amsmath}
\usepackage{graphicx}
\usepackage{dcolumn}
\usepackage{bm}
\usepackage{alltt}%
\usepackage[normalem]{ulem}
\usepackage{natbib}
\usepackage{adjustbox}

\newcommand{\mS}{\mathcal{S}}

\begin{document}

\author{Bradley M. Dickson}
\email{bradley.dickson@vai.org}
\affiliation{Center for Epigenetics, Van Andel Research Institute, Grand Rapids Michigan 49503, USA.}
\author{Parker W de Waal}
\affiliation{Center for Cancer and Cell Biology, Van Andel Research Institute, Grand Rapids Michigan 49503, USA.}
\author{Zachary H Ramjan}
\affiliation{Center for Epigenetics, Van Andel Research Institute, Grand Rapids Michigan 49503, USA.}
\author{H Eric Xu}
\affiliation{Center for Cancer and Cell Biology, Van Andel Research Institute, Grand Rapids Michigan 49503, USA.}
\author{Scott B. Rothbart}
\affiliation{Center for Epigenetics, Van Andel Research Institute, Grand Rapids Michigan 49503, USA.}
\email{scott.rothbart@vai.org}

\title{A fast, open source implementation of adaptive biasing potentials uncovers a ligand design strategy for the chromatin regulator BRD4}

\begin{abstract}
In this communication we introduce an efficient implementation of adaptive biasing that greatly improves the speed of free energy computation in molecular dynamics simulations. We investigated the use of accelerated simulations to inform on compound design using a recently reported and clinically relevant inhibitor of the chromatin regulator BRD4. Benchmarking on our local compute cluster, our implementation achieves up to 2.5 times more force calls per day than plumed2. Results of five 1$\mu$second-long simulations are presented, which reveal a conformational switch in the BRD4 inhibitor between a binding competent and incompetent state. Stabilization of the switch led to a -3 kcal/mol improvement of absolute binding free energy. These studies suggest an unexplored ligand design principle and offer new actionable hypotheses for medicinal chemistry efforts against this druggable epigenetic target class.
\end{abstract}

\maketitle
\section{Introduction}
Molecular dynamics (MD) simulations are often invoked within the context of drug discovery as a means to score ligands by their absolute binding affinity.\cite{dill} In practice, MD simulations are also very useful in the discovery process if they contribute testable hypotheses that guide exploration of the chemical space around a ligand.\cite{camerino,stuckey} To be influential, simulations need to provide actionable information on a short time scale, such that a team of medicinal chemists cannot make the same discovery empirically or find an alternative route to success before simulation results are in hand. 

\begin{figure}
\includegraphics[width=\columnwidth]{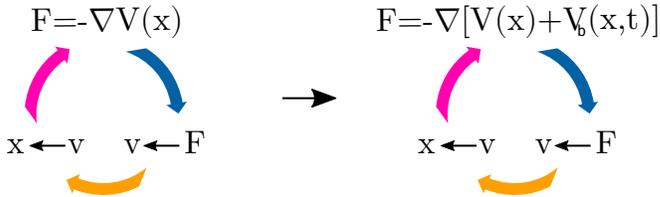}
\caption{A basic schematic of the MD propagation cycle before (left) and after (right) the addition of an adaptive biasing potential. The number of trips around the cycle per unit time expresses the speed of MD simulation.}\label{cycle}
\end{figure}

The ability to explore the binding process and to compute the free energy of binding within a relevant time frame {\it in silico} depends on the rate at which the governing equations of motion can be propagated (Figure \ref{cycle}). Standard MD is impractical for describing ligand-protein binding/unbinding because more than $0.5\times 10^{9}$ force evaluations must be made before the simulation reaches microsecond timescales. Meanwhile, effective ligands may stay bound to their targets for tens of minutes\cite{rates}. With commonly used simulation tools, it takes about a week to simulate a microsecond of a protein-ligand system (Figure \ref{timesMPI}). Therefore, a considerable gap exists between the rate at which one can perform ligand binding simulations and the rate at which empirically driven progress can be made in a drug discovery setting.

A recent  approach to tackle this problem of timescales has been the construction of very specialized computers\cite{anton,anton2}, which has generated some long trajectories of ligand binding\cite{shawbinding,shaw2}. These specialized computers aim to make as many MD cycles per unit time as possible. An alternative, more easily dispersed approach is the use of algorithms that leverage formalities of statistical mechanics to extract as much information as possible from every iteration of the MD cycle. This approach aims to reduce the number of cycles required to estimate thermodynamic quantities such as free energy of binding.

In this study, we focus on the use of adaptive biasing potentials (ABP) to increase the efficiency of free energy calculations and conformational exploration. ABP methods supplement the system energy in a time-dependent manner so that metastable states are slowly flooded with energy until they are flattened away. The flooding leads to new forces that are contributed by a time evolving biasing potential $V_b$ (Figure \ref{cycle}). 

In the MD cycle depicted in Figure \ref{cycle}, ABP methods introduce computational overhead in evaluation of the biasing force, and thus slow the rate of completing the MD cycle. While native MD engines are typically faster at making cycles, ABP schemes improve the statistical value of each cycle by several orders of magnitude. The exact improvement of statistical efficiency can be expressed as the ratio of unbiased to biased partition functions, which quantifies exactly how the ABP schemes improve sampling efficiency by shrinking the Boltzmann-weighted volume of configuration space.

Given the statistical advantage brought about by ABP, the inefficiency of the associated MD cycle is often taken for granted. Here we show that by minimizing ABP network communications within the MD engine and by minimizing computational complexity of the ABP scheme, the efficiency of the biased MD cycle can be drastically improved while maintaining the statistical advantage of ABP. On our local compute cluster, we show efficiency gains of 25\% to 67\% over the commonly used plumed2\cite{plum2} plugin. In Amazon's elastic cloud computing environment, we demonstrate a 36\% improvement in the MD cycle rate for a 26\% reduction of simulation costs compared to the same plumed2 plugin. These improvements reduce the time gap between {\it in silico} and {\it in vitro} drug discovery, and increase the affordability of high throughput cloud applications. In light of the expanding ABP usage base, these efficiency gains can also be projected to generate substantial savings and an improvement in simulation scope for the computational community. 

As an example application, the results of 4 microsecond-long simulations of a ligand-protein system are reported for a clinically relevant inhibitor\cite{toy} against the chromatin regulator BRD4 (bromodomain-containing protein 4). Three of these simulations were allowed to sample the full surface of the protein and 60\% of the simulation volume to fully detail the protein-ligand interaction. These simulations show several metastable states where the ligand aggregates on the protein surface and suggest that better control over a ligand conformation change could improve affinity. The current ligand scaffolds present two conformations, only one of which is binding competent. 

Without the efficiency gains derived below, each one of the $\mu$s-long simulations presented in this work would have taken 5.3 days longer to complete on exactly the same computer hardware had the current gold standard ABP software been used. Had all of the simulations been run in serial, they would have taken an extra 21 days to complete. In Amazon's elastic cloud computing environment, the simulations presented below would cost an additional $\$1200$ USD without the time-saving advances described here. Extrapolating these savings to the MD community projects years of saved compute time and tens of thousands of dollars. Thus, the implementation described below positively impacts computational throughput and better optimizes the dollars spent funding computational research. 

\section{Results and Discussion}
The efficiency of the ABP implementation has been improved by considering two sources of computational overhead: (1) {\it Computational complexity of the bias potential.} In our implementation, the master node of the simulation handles the bias updates and bias force evaluation which is a concern for load balancing. Thus, a major advance in this implementation model has been the reduction of computational complexity within the bias itself. We have written the bias in terms of kernels, or ``hills,'' that have compact support (Figure S1). These new hills result in a significant reduction of operations (namely reduced memory accesses and multiplies) when the bias is updated, leading to better load balancing. Others have also used truncated kernels to reduce complexity\cite{babin} and below we demonstrate that this one aspect of the implementation is not enough to ensure good scalability. (2) {\it Minimization of communications introduced by the ABP scheme.} In ABP schemes, a collective variable (CV) must be computed. The CV is a coarse-grained description of the system state, and as such, it requires atomic positions. In most modern MD engines, the master node does not know all of the current atomic positions, so some communication is required. Additionally, once the master node computes the biasing force, that force must be communicated back to the nodes that are managing atoms in the CV. Details regarding each of these aspects of development are given in the SI.

We implemented two bias potentials, with the default being mollified adaptive biasing potential (mABP)\cite{pre11}. This bias potential was derived elsewhere\cite{pre11,mu15} and has been shown to recover well-tempered metadynamics\cite{bbp08} under an approximation that destroys the mollifying properties of the bias\cite{mu15}. Importantly, mABP is not a form of metadynamics. Reference \onlinecite{mu15} shows that while mABP defines the bias potential as the log of a histogram of sampling, all forms of metadynamics write the bias potential as directly proportional to a histogram of sampling. Following from this, a key attribute of mABP is that the biasing force is an ensemble average quantity. The metadynamics family tend toward higher computational complexity than mABP. Boundary corrections\cite{boundary,metabasin}, for example, are not required in mABP and are not supported in this distribution. Convergence behavior for mABP and well-tempered metadynamics (WTmetaD) are reported in Supplemental Material for alanine dipeptide to test functionality with the compact hills (Figure S2). 

It is known that metadynamics places strict conditions on the hill functions\cite{dama2014} and that the convergence time of metadynamics scales exponentially with the hill width\cite{mu15}. We observed a parameter dependent sensitivity to artifacts for the well-tempered metadynamics method when using the hill functions introduced in this work (Figure S3). Artifacts could be observed when WTmetaD was used to repeat the simulations described below as well, though the artifacts were much more difficult to diagnose. Therefore, we do not recommend invoking the metadynamics method with the hills implemented here. WTmetaD performs well for some hill settings but not for others, leaving open the possibility of undiagnosed failures when using the method. 

\subsection{Timing of the implementation for a practical application}
\begin{figure}
\includegraphics[width=\columnwidth]{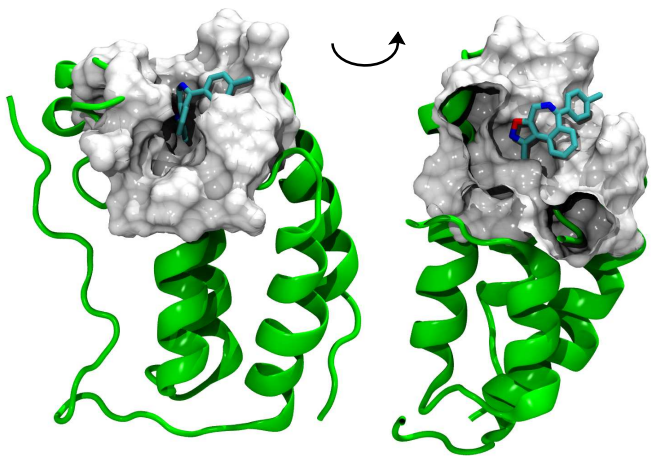}
\caption{(Left) PDB entry 5hm0 is shown with the ligand in green sticks. (Right) The 5hm0 entry is shown again where a clipping plane is used to expose the entire ligand.}\label{system}
\end{figure}

 An application to ligand binding is given here to measure overhead of the new implementation and as a demonstration of the mABP method. The model system involves a recently reported, clinically relevant, small molecule which binds the first bromodomain of BRD4\cite{toy} (PDB:5hm0). BRD4 is an acetyllysine binding protein that is emerging as a therapeutic target for controlling expression of the ``undruggable'' oncogenic transcription factor MYC\cite{firstdrug}. Inhibitors like the one simulated here are part of an emerging class of epigenetic-based therapeutics being tested clinically for the treatment of haematological malignancies.\cite{betreview} The lead ligand (compound 3 of reference \onlinecite{toy}) in complex with bromodomain 1 of BRD4 is shown in Figure \ref{system}. 

A pair of RMSD CVs were chosen for the simulations. One CV is the RMSD between a subset of atoms in the ligand at its current (simulated) position and the Xray structure. The second CV is the RMSD between a different subset of ligand atoms in its current (simulated) position and the Xray. This set of CVs projects the full ligand-accessible space onto the diagonal of the CV space because the two CVs are tightly coupled (Figure S4). The base GROMACS input topologies and tpr files for equilibration were built according to the swissparam tutorial for preparing GROMACS topologies\cite{swissparam}. Everything required to re-run these biased simulations (all simulation parameters and inputs), including a general tutorial for using the ABP code, can be found at the GitHub page for this project.\cite{megit2} The tutorial itself allows one to perform simulations of this ligand-protein system.

\begin{figure}
\includegraphics[width=\columnwidth]{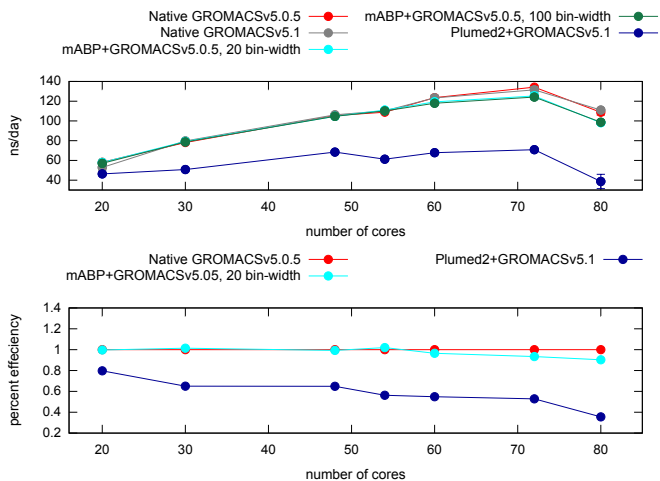}
\caption{(Top) Nanoseconds per day as a function of compute cores for GROMACS, mABP (fABMACS), and plumed2. (Bottom) Percent efficiency relative to native GROMACS as a function of cores. All data points are an average of three runs.}\label{timesMPI}
\end{figure}

The performance of our ABP implementation is compared against native GROMACSv5.0.5 and native GROMACSv5.1 for several numbers of compute cores. Because the implementation builds on the GROMACS package, we call it fABMACS for \uline{f}ast \uline{A}daptively \uline{B}iased \uline{MA}chine for \uline{C}hemical \uline{S}imulations. The data shown in Figure \ref{timesMPI} were selected to minimize oscillations in the scaling of the native GROMACS codes, representing the envelope of best performance for the native GROMACS MD engine. In all cases, GROMACS was allowed to choose the partitioning of PME and particle-particle nodes and in all cases we verified that all the data at a particular node count were collected with the same partitioning. For the fABMACS simulations, the CV space was discretized into a 480 by 480 bin grid with an RMSD range from 0 to 12 nanometers. Although irrelevant for timing, the biasing parameters were $c=0.001$, $b=0.9$. A range of hill widths were considered, and the temperature was 300 Kelvin. Results are reported in units of nanonseconds per day, which can be converted to the number of MD cycles per day by multiplying by 500000 (the MD timestep was 0.002 fs). See SI section 1 for the definition of the hill functions, $\mS_a^p$. The parameter $a$ fixes the width of the hill and $p$ alters the shape. Section 3 of the SI covers how to think about these hills in terms of the more common Gaussian hill.

Figure \ref{timesMPI} also shows the performance of the plumed2\cite{plum2} plugin running WTmetaD, which was plugged in to GROMACSv5.1. The fABMACS parameters were converted to well-tempered metadynamics parameters according to reference \cite{mu15} and the hill width used with plumed was within the range of widths shown for fABMACS. The CV and CV discretizations were identical for the plumed2 and fABMACS methods. All of the various GROMACS codes were compiled with the same configurations on the same hardware. We did not observe inefficiency when running WTmetaD in fABMACS. WTmetaD introduces three extra multiplies per grid point during the bias update (to scale hill heights for the potential and its forces), and one exponential evaluation (to determine the hill height). On 72 cores, our implementation of WTmetaD (without boundary corrections) averaged only 1 ns/day less than mABP. 

Our ABP implementation presents minimal overhead to the underlying GROMACS engine, and the scalability of GROMACS is intact. For this system, fABMACS achieves a maximum of 125 ns/day on 72 compute cores spanning three different machines. The maximum of the native GROMACS code is 134 ns/day on 72 cores. It can be appreciated, as expected, that the extra ABP-related communications are slowing down the simulation as the number of cores is increased (Figure \ref{timesMPI}; bottom). 

Given the rise of cloud computing, it is pertinent to consider timings in the cloud environment as well. GROMACSv5.1, plumed2, and fABMACS were timed in Amazon's EC2 environment using ``g2.8xlarge'' compute nodes. These nodes have 32 cores and four Nvidia GRID GPUs. The cost of this EC2 instance is currently $\$2.60$ USD per hour. Table \ref{tab1} shows the cost per microsecond for native GROMACS, fABMACS, and plumed2 based on timings of 100000 MD steps. Note that both plumed2 and fABMACS should be expected to generate better sampling than native GROMACS, due to the shrinking partition function, even though native GROMACS is the least expensive per force call. Running fABMACS costs \$$38$ USD more per microsecond than native GROMACS, while plumed2 costs an extra \$$326$ USD per $\mu$s. We did not run simulations spanning multiple machines in the cloud because projected costs did not suggest any benefit.

\begin{table}
\caption{Amazon Elastic Compute Cloud Results, and cost in USD/$\mu$s and days/$\mu$s.}
\label{tab1}
\def\arraystretch{1.5}
\begin{adjustbox}{max width=\columnwidth}
\begin{tabular}{|c|c|c|c|c|}
\hline
Software & Biased & ns/day & $\$(\text{USD})/\mu\text{s}$ & Days/$\mu$s\\
\hline
GROMACSv5.1 & no & 82.027 & \$760 & 12.2\\
fABMACS & yes & 78.136 & \$798 & 12.8\\
plumed2 & yes & 57.449 & \$1086 & 17.4\\
\hline
\end{tabular}
\end{adjustbox}
\end{table}

\begin{figure}
\includegraphics[width=5in]{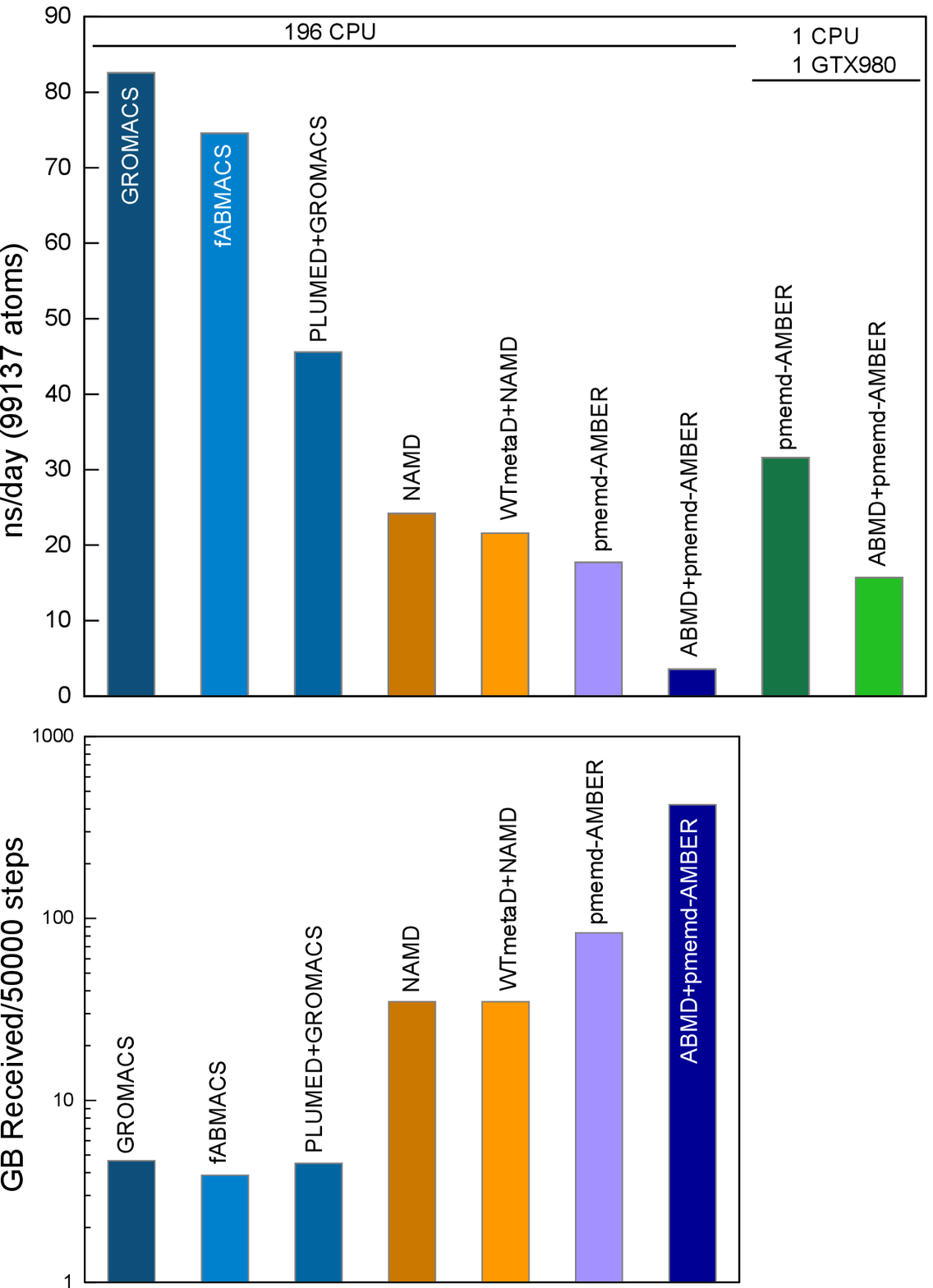}
\caption{Universal benchmark using 99,137 atoms for adaptive biasing of two 4-atom CVs, in the canonical ensemble with partical mesh Ewald and periodic boundaries.\cite{bench} (TOP) Nanoseconds per day for different MD packages on CPU or GPU. (BOTTOM) Network traffic on a node for MD packages when distributed across 7 compute nodes.}\label{universal}
\end{figure}
Encouraged by the performance of the application tested above, we also tested the performance of the built-in NAMDv2.11\cite{NAMDref} and AMBERv16.05\cite{amberref,amber2} implementations of adaptive biasing potentials for a larger universal benchmark. The benchmark system is a small peptide in water, producing a 99k atom system.\cite{bench} All collective variables were 2 RMSDs to an initial position with each CV having four atoms. Figure \ref{universal} reports results similar to what others have observed for unbiased simulations.\cite{comp2006,gromacs4} Here, we have used 28 cores spread across 7 nodes and to gain insight to the performance of the different MD packages we watched the network communications on the master node of the simulations via ifconfig. The cumulative inbound traffic is shown in Figure \ref{universal}(Bottom) for each MD package. All machine interconnects were 40GbE. 

Figure \ref{universal}(Bottom) suggests that PLUMED2 is inefficient within the bias update and bias force computations but not in network communications. This may be a load balance issue. The built-in NAMD biasing seems limited by the underlying speed of NAMD. The pmemd-AMBER implementation appears to produce excessive network traffic when ABMD is invoked. Activation of biasing within AMBER on GPU also incurs efficiency losses either because excessive system information is passed from the GPU to CPU for biasing at every timestep or because CPU speed is a bottleneck. As we noted earlier, this demonstrates that minimization of operational complexity in the bias\cite{babin} is insufficient to achieve good scalability although it is required for load balancing.


\begin{figure*}
\includegraphics[width=6in]{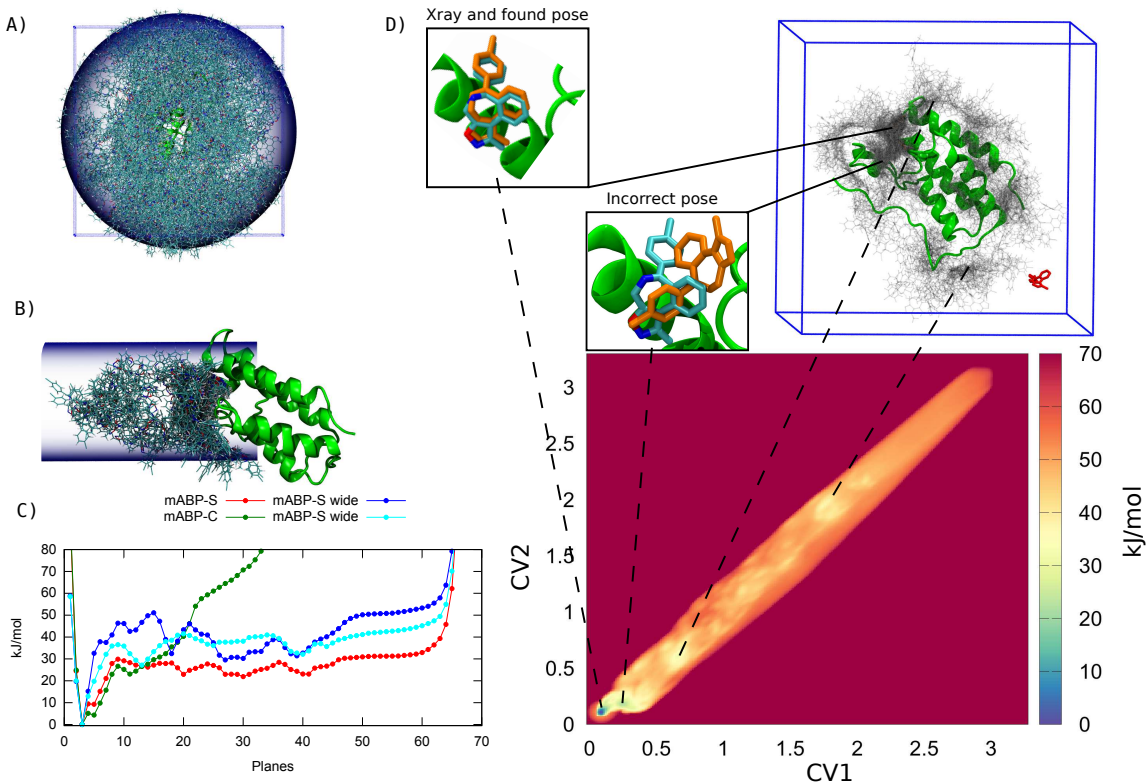}
\caption{(A) Snapshots of the ligand from a 1$\mu$s simulation with a spherical restraint. The simulation box is shown in blue. (B) Snapshots of the ligand from a 1$\mu$s simulation with a cylindrical restraint. (C) 1D projections of the free energy onto CV$_1$. The ``-S'' indicates that the spherical restraint was used while ``-C'' indicates that the cylindrical restraint was used. The narrow and wide designations are described in the main text. (D) Snapshots of the ligand where at least one ligand atom is within 3.5 Angstrom of the protein, highlighting some features of the landscape. Simulated ligand poses are shown in orange in the inset panels. Initial ligand position is shown in red.}\label{res1}
\end{figure*}

\subsection{Applications to ligand binding}
The fABMACS package can be compiled for different scenarios. To keep the code as streamlined as possible, a number of options are hard-coded via a patching script. The script sets a number of parameters for things like CV space discretization, selection of mABP or WTmetaD, and what type of restraint system to apply. There are two selectable restraint types, a spherical or cylindrical restraint. There is a third restraint that cannot be deselected, which limits the maximum value of the CVs. This restraint can be obsoleted by setting a large maximum limit. A full tutorial is available on the project GitHub page\cite{megit2}, which walks through setting up and running the simulations presented in this section. The periodic simulation boundaries are unwrapped such that the ligand is whole and such that the ligand does not ``jump'' across the box. Unwrapping is not supported in some ABP plugins\cite{pbcplumed}, yet unwrapping is critical unless the simulation system is designed such that the CVs are never broken by periodicity. 

We performed a total of four 1$\mu$s simulations with the lead ligand that used the spherical (Figure \ref{res1}A) or cylindrical (Figure \ref{res1}B) restraint. The mABP simulations used hills with compact support , called $\mS_a^p$, where $p=20$ and $a=10\times \Delta \xi_i$ or $p=20$ and $a=20\times \Delta \xi_i$, where the latter is labeled ``wide'' in Figure \ref{res1}D. Plots of free energy for all of the simulations are shown in 2-dimensions (Figure S4) and as 1-dimensional projections (Figure \ref{res1}D). We followed the formalisms of References \onlinecite{Funnelex1} and \onlinecite{acemd} for estimating absolute binding free energy with and without restraints, respectively.

The absolute free energy of binding estimated from each of the simulations ranges from -5.18 to -7.63 kcal/mol. All of the estimates are within 4 $k_BT$ of one another, where $k_B$ is Boltzmann's constant and $T$ is temperature. This is a surprising degree of similarity given that three of the simulations were allowed to sample the entire protein surface, and two of those started in an unbound conformation. In each of the simulations using wide hills and the large spherical restraint, the crystallographic ligand pose was found only once. One of these simulations started in the bound conformation, one started in an unbound conformation. Both simulations found the crystallographic pose from an unbound state. In the simulation using narrow hills and starting in an unbound conformation, the crystallographic ligand pose was found twice. (Figure S6) Simulations as loosely restrained as these are very rarely performed and have even been described as impossible.\cite{Funnelex1} Indeed, we only tried these simulations after being encouraged by the performance of fABMACS. The simulation using the cylindrical restraint found the crystallographic pose four times. Thus, the free energy estimate from the cylindrical restraint is expected to be most accurate.

The average absolute free energy of binding over the four estimates is -6.56 kcal/mol with a standard error of 0.56 kcal/mol. The free energy estimate from simulation with the cylindrical restraint was -6.16 kcal/mol. We did not compare to binding free energy estimated from IC$_{50}$ reported in Reference \onlinecite{toy} for two reasons. First, we did not parameterize the ligand via {\it ab initio} methods. Second, the binding constant of the competitor ligand must be known to make an accurate free energy estimate\cite{noic50}, and details of the competitor ligand were not reported. 

It is also worth noting that most ligand development efforts focus on dissociation constants rather than free energy of binding. If simulation estimates leave a 1 kcal/mol error in the free energy, there is a 5-fold error in the dissociation constant. For a 2 kcal/mol error in the free energy, there is a 29-fold error in the dissociation constant. In general, it is more informative to generate structure-activity trends than to compare simulation results to experimental metrics. 

The simulations using the spherical restraint sample a number of configurations where the ligand is aggregated to the protein surface (Figure \ref{res1}C). We observe a large number of states where the ligand is aggregated on the ZA-loop or docked in the binding pocket with an incorrect pose. Most binding pathways show the ligand aggregate to the pocket or ZA-loop and then explore different ligand poses and azepine ring conformations until the crystallographic pose is found. Often, the ligand leaves the pocket area and moves back into solvent before finding the correct pose.

Many attempted binding events were rejected by the protein because the ligand adopted a binding incompetent conformation. The benzoisoxazoloazepine ligand contains an azepine ring that can adopt different puckering states; (Figure S5) only one of the two major puckering states is consistent with the crystallographic pose. The ``incorrect pose'' shown in Figure \ref{res1}C is representative of a state in which the azepine ring has adopted a conformation inconsistent with the crystallographic pose, precluding correct binding and slowing the binding process. 

\begin{figure}
\includegraphics[width=\columnwidth]{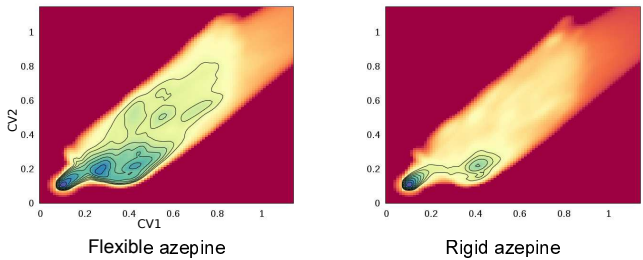}
\caption{Binding landscapes without (left) and with (right) a rigid azepine. Contours are every $2k_BT$ from 0 to 40 kJ/mol.}\label{nonflex}
\end{figure}

As a simple means of testing the hypothesis that binding could be improved by reducing flexibility of the azepine ring, we computed the binding free energy again while applying a dihedral restraint that prevents the azepine from changing conformations. The absolute binding free energy, computed from a 1$\mu$s simulation using the cylindrical restraint and an inflexible azepine, was -9.42 kcal/mol. This compares to -6.16 kcal/mol for the flexible azepine in the cylindrical restraint (Figure \ref{nonflex}). The restrained ligand found the crystallographic pose 6 times (Figure S6). These simulations validate our hypothesis and motivate ligand chemistry that would restrain the azepine. Functionalization of the ligand or abandonment of the azepine ring are both plausible directions. 

\section{Conclusion}
We have introduced and benchmarked an implementation of ABP in GROMACS 5.0.5, called fABMACS, that presents minimal overhead to the GROMACS engine. The implementation strives to maintain the scalability of GROMACS. A minimal communication scheme and local smoothing kernels have been used to achieve this goal. Drastic efficiency gains were demonstrated, as compared to the most common alternative ABP plugin.

The fABMACS code is available on GitHub\cite{megit2} where a tutorial can also be found. All of the ligand-protein simulations presented above can be re-run by following the tutorial. Currently, only RMSD CVs are supported, but a range of systems should be compatible with this. A future goal is to develop the distribution to support a range of different CVs. Cylindrical, spherical, and simple range-limiting CV restraints are supported. A build script is included with the code, which customizes source code for user specifications of methods, restraint types, restraint scales, and so-on. Thus, if a user can install GROMACS 5.0.5 on their compute infrastructure, then fABMACS can be installed using the same configuration. Simply run the build script with specified options, and compile the source code as normal. Implementation into other MD engines are also ongoing.

The results of long simulations of the first bromodomain of BRD4 with a clinically-relevant small molecule were presented, both to make methodological demonstrations using mABP and to examplify structural insights that are obtained from atomistic simulations. Simulations were presented that found the bound state after having escaped, even when the full surface of the protein could be sampled. We suggest a trend in the published SAR data that could be further exploited to improve potency --- Namely the azepine ring affords conformational freedom that reduces ligand efficiency.  

fABMACS will be useful in identification of false positives, creation of chemical hypotheses, scoring ligands, and guiding NMR based fragment screening. The efficiency gains of fABMACS will enable more efficient use of computational resources and make new, ambitious applications more affordable and tractable.

\section{Supplemental Material: A fast, open source implementation of adaptive biasing potentials uncovers a ligand design strategy for the chromatin regulator BRD4}

\subsection{Streamlined bias potential updates}
Let the CV be $\bm\xi$ which is a vector of coordinates in the CV space, and let $\bm\xi^*$ be a particular point. We use $\bm x_t$ to indicate the position of the trajectory in configuration space, and $\bm\xi(\bm x_t)$ or $\bm\xi_t$ to indicate the trajecotry position in CV space. The mABP bias potential is 
\begin{equation}\label{apfm}
V_b(\bm{\xi},t) = \beta^{-1} \frac{b}{1-b}\ln[c\,(1-b)\,\displaystyle h_{\alpha}(\bm{\xi},t) +1]
\end{equation} and the mABP bias gradient is 
\begin{equation}\label{btraj-grad}
\frac{\partial V_b(\bm{\xi}(\bm x_t),t)}{\partial \xi_i} = 
\frac{c\, b\, \beta^{-1}\,h_{i,\alpha}'(\bm{\xi}^*,t)}
{1+c\,(1-b)\,h_{\alpha}(\bm{\xi}^*,t)} 
\end{equation} where $\beta^{-1}=k_{B}T$ is the temperature of the simulation in units of energy. $b$ and $c$ are parameters, as discussed in the Appendix (or SI). Two histograms are required for evaluating the biasing force
\begin{equation}\label{histos}
\begin{split}
&h_{\alpha}(\bm{\xi},t) = \displaystyle\int_0^t\delta_{\alpha}(\bm{\xi}(\bm x_s)-\bm{\xi})\, ds\\
&h_{i,\alpha}'(\bm\xi,t) = \displaystyle\int_0^t \partial_{\xi_i}\delta_{\alpha}(\bm\xi(\bm x_s)-\bm\xi)\, ds 
\end{split}
\end{equation} The first is a standard histogram convoluted with a smoothing kernel, or ``hill'', 
\begin{equation}\label{ghill}
\delta_{\alpha}(\bm{\xi})=\exp\left(-\frac{|\bm{\xi}|^2}{\alpha^2}\right)
\end{equation} where $\alpha$ controls the width of the Gaussian. The second histogram is actually the weak derivative of the raw (un-smoothed) histogram along $\bm\xi$. One can appreciate that if the histograms are collected, the biasing force requires only a look-up in the histograms and some arithmetic operations. As we have advocated in the past\cite{ens10}, this is also much more simple than adaptive biasing force methods\cite{dp01,dp08}.

In order to collect these histograms with computer code, the CV (and time) must be discretized. Let the discretization use $Nbins$ for each dimension of the CV space. Ignoring the trivial details, the position of the trajectory is then approximated by the nearst grid point $\bm\xi^*\approx \bm\xi(\bm x_t)$ and the histograms are updated as
\begin{equation}\label{update}
\begin{split}
&h_{\alpha}(\bm{\xi},t+\Delta t) = h_{\alpha}(\bm{\xi},t)+\delta_{\alpha}(\bm{\xi}^*-\bm{\xi})\\
&h_{i,\alpha}'(\bm\xi,t+\Delta t) = h_{i,\alpha}'(\bm{\xi},t)+\partial_{\xi_i}\delta_{\alpha}(\bm\xi^*-\bm\xi) 
\end{split}
\end{equation} 
for all grid points $\bm\xi$ at every timestep of the simulation with the initial conditions $h(\cdot,t=0)=0$ and $h'(\cdot,t=0)=0$. ($\Delta t$ is the MD timestep) The hills (or smoothing kernels) are stored in a lookup table (see reference \onlinecite{mu15}) by precomputing $\delta_{\alpha}(\bm\xi^*-\bm\xi)$ for every possible pair $(\bm\xi^*,\bm\xi)$ in the discretized CV space. Thus the Gaussians $\delta_{\alpha}$ are evaluated once and only once. However, because the Gaussian hills do not vanish, the update in equation \eqref{update} is made for all grid points. In order to greatly reduce the number of arithmetic operations in equation \eqref{update}, one can introduce smoothing kernels that have compact support. Such kernels would only be non-zero in the neighborhod of $\bm\xi^*$ so that an update only needs to cover a small subset of the grid points representing the CV space.

For each component of the CV $\xi_i$, the compact smoothing kernels have the form (see figure \ref{showhill})
\begin{equation}\label{compacthill}
\mS_a^p(\xi_i) = \begin{cases}
e^p\times e^{\left(\frac{-p}{1-|\frac{\xi_i}{a}|^2}\right)} \text{if $|\frac{\xi_i}{a}|<1$}\\
0, \text{otherwise}
\end{cases}
\end{equation} where the leading $p$ factors of $e$ ensure that $\mS_a^p(\xi_i=0)=1$ and the parameter $a$ controls the width of the function. To further minimize complexity, we define the smoothing kernel in a multi-dimsional CV space as the product of these one dimentional functions. The integer $1\leq p$ allows the resulting multidimensional function to be round rather than square. (Increasng $p$ makes the functions more round.) With these compact functions the update becomes 
\begin{equation}\label{Cupdate}
\begin{split}
&h_{a}(\bm{\xi},t+\Delta t) = h_{a}(\bm{\xi},t)+\mS_{a}^p(\bm{\xi}^*-\bm{\xi})\\
&h_{i,a}'(\bm\xi,t+\Delta t) = h_{i,a}'(\bm{\xi},t)+\partial_{\xi_i}\mS_{a}^p(\bm\xi^*-\bm\xi)
\end{split}
\end{equation} 
where the update only runs over grid points near $\bm\xi^*$, such that $\mS_{a}^p(\bm\xi^*-\bm\xi)>0$, {\it i.e.,} those grid points within a distance $a$ from the trajectory at $\bm\xi^*$. The WTmetaD update can be written in analogy with the mABP update, with the only difference being that WTmetaD scales the update in equation \eqref{Cupdate} by a factor determining the so called ``hill-height''. The value $a$ may be different for different components of $\bm\xi$ although this case is not covered here. For a multidimensional CV space this results in a large reduction of evaluations at each bias update. In two dimensions, we replace $Nbin^2$ operations with $Nbase^2$ operations where $Nbase$ is the number of grid points spanned by $\mS_a^p$. All the analysis of mABP from reference \onlinecite{pre11} holds with these new functions so the method will converge, as we next demonstrate.

\subsection{Minimization of Interprocess Communications}
To minimize communications, each processor checks if it is responsible for any atoms in the CV. If a processor does manage CV-implicated atoms, it sends a flag to the master processor to indicate this, and it sends a vector to the master which holds those atom's coordinates. If a processor does not manage atoms in the CV, then it sends a flag to indicate this. The master now knows which processors are involved in CV and which processors are not, and recieves the atom coordinates from those processors that are involved in the CV. The master then passes the atomic coordinates (of only those atoms in the CV) to the mABP routine which computes the bias update and bias force. The master then sends the bias force back to the processors that have atoms involved in the CV, where each processor then adds the biasing force to the GROMACS force for its atoms. 

This communication scheme was implemented by injecting MPI (message passing interface) function calls in the md.c routine of GROMACS. The GROMACS ga2la\_get function was used to determine which atoms are ``home'' atoms for each process. The ``home'' atoms that are in the CV are the atoms that must be sent to the master for adaptive biasing. All of the additional code, save initializations, was injected after the call to do\_force. Our biasing routine is written in fortran and named ``hellof'', meaning ``hello fortran.'' The biasing function as well as the fortran compilation option were specified in the CMakeLists.txt file of the mdrun directory. No other changes to GROMACS were required.

\subsection{Convergence with $\mS_a^p$}
The usual alanine dipeptide benchmark is used here to demonstrate convergence for this new smoothing function. Exactly the same set-up for measuring convergence, the same reference free energy, and the same biasing parameters $(c\,\text{and}\,b)$ are used as recently reported\cite{mu15}. 

\begin{figure}[h]
\includegraphics[width=\columnwidth]{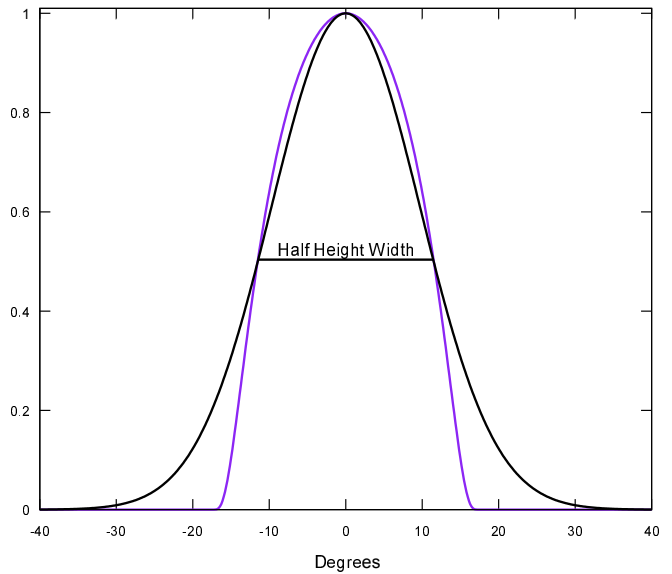}
\caption{Compact hill (purple) with a Gaussian hill (black).}\label{showhill}
\end{figure}

The CV space here is two dimensional, and each component is a backbone dihedral angle, one CV for $\psi$ and one for $\phi$. In Figure \ref{showhill}, we plot $\mS_{18}^1(\xi_1)$, where $\xi_1$ is one of the dihedral angles. Using $a=18$ scales the domain of $\mS_a^1$ so that the kernel is zero when the dihedral angle is more than $18$ degrees from $0$. The compact hill $\mS_a^1$ is shown in purple, while the more typical Gaussian hill is shown in black. The width of the two hills at half-height is identical, and we suggest that the width at half-height is a reasonable way to get some intuition for these smoothing kernels as compared to the Gaussian kernels. 

The width at half-height for $\mS_a^p$ is given by 
\begin{equation}
w_a = 2\,a\,\sqrt{\frac{\ln(2)}{p+\ln(2)}}
\end{equation} The base of $\mS_a^p$ has a width of $2a$, and a bin-width of $Nbase = 2a/\Delta\xi_i$ where $\Delta \xi_i$ is the distance between grid points for the $i$-th CV. For equation \eqref{ghill}, one can write the $\alpha$ for which $\delta_{\alpha}$ has the same width at half-height $\alpha = w_a/(2\sqrt{\ln(2)})$. This allows one to have a feeling for how $\mS_a^p$ compares to a more familiar Gaussian smoothing kernel. The Gaussian in Figure \ref{showhill} uses $\alpha = 13.833$.

\begin{figure}[h]
\includegraphics[width=\columnwidth]{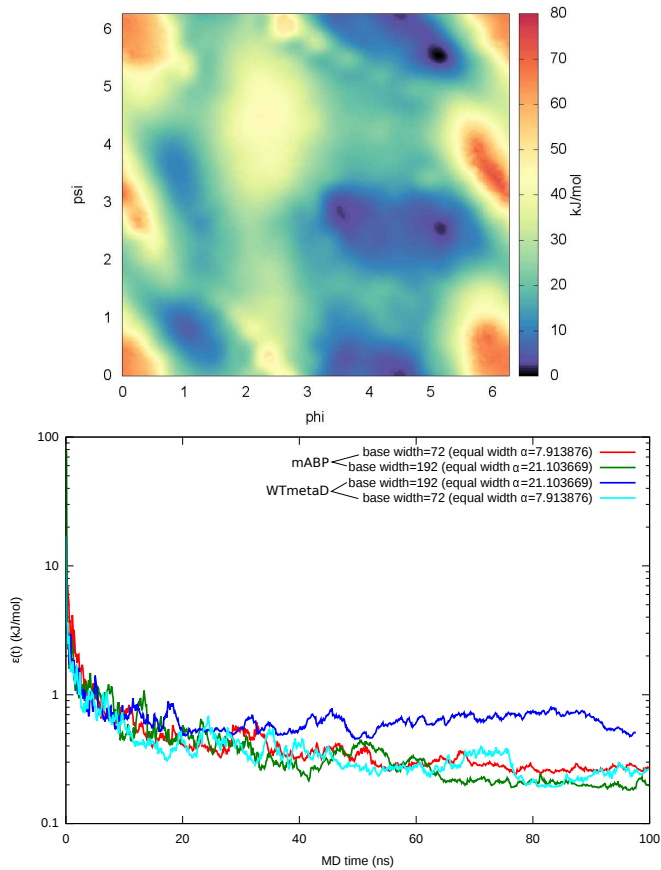}
\caption{(Top) Free energy landscape from mABP where the base-width of $\mS_a^{p=1}$ was $72$ degrees. (Bottom) Convergence of alanine dipeptide with compact hills for different base widths ($2a$) in units of degrees. (all $p=20$)}\label{converg}
\end{figure}

In Figure \ref{converg}, error in the free energy estimate is shown as a function of simulation time for different values of $a$. The error is the average absolute difference between the ABP (either mABP or WTmetaD) free energy estimate and a reference free energy, as described in reference \onlinecite{mu15} One of the strengths of mABP was that it leads to a free energy estimate that is independent of $\alpha$.\cite{pre11} When using these new kernels, the free energy is independent of $a$, as can be seen in Figure \ref{converg}. WTmetaD, on the other hand, seems to be more sensitive to these hills than it is to Gaussian hills. Figure \ref{converg} reflects this, showing an impaired convergence for a hill that compares to a Gaussian of about 20 degrees. We've recently looked in detail at how metadynamics is impaired when the hill widths are large\cite{mu15}, where we showed WTmetaD converges quickly until the Gaussian hill widths are greater than 30 degrees. 

\begin{figure}[h]
\includegraphics[width=\columnwidth]{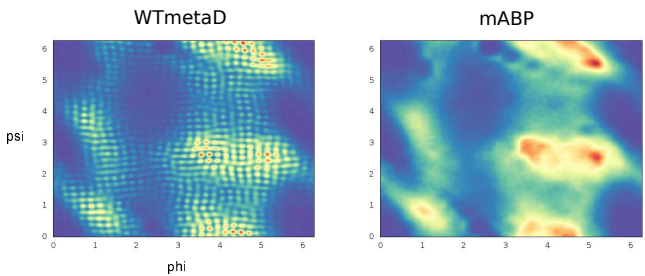}
\caption{Histograms of the trajectory sampling for WTmetaD and mABP with the same hill settings. WTmetaD shows a corrugated pattern while mABP shows a smoother sampling.}\label{hists}
\end{figure}

The $\mS_a^p$ hills have a boxy shape in 2D. The parameter $p$ allows them to be rounded out. The results in Figure \ref{converg} used $p=20$. The problem for WTmetaD is that we do not know how to set $p$ and $a$ for an arbitrary system in a way that will guarantee sufficiently round hills. The WTmetaD method demonstrates a failure to converge on the timescales we tested when $p=1$ and it demonstrates pathological sampling behaviors. Figure \ref{hists} shows the histogram of backbone dihedral angles that was collected from WTmetaD and mABP using the same hills (base width $24$ degrees and $p=1$). While it is possible to choose functional hills for WTmetaD, it is impossible to know {\it a priori} which hills will fail. mABP, on the other hand, exhibits robust performace. We stress that increasing $p$ salvages the WTmetaD simulations but a general rule for identifying failure is not in-hand. In the mean time, we advocate sticking to mABP while using fABMACS.


\subsection{Ligand Binding}

\begin{figure}[h]
\includegraphics[width=\columnwidth]{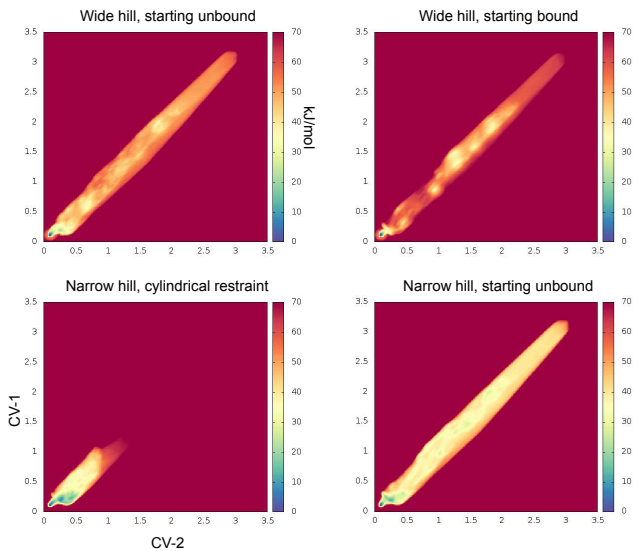}
\caption{Landscapes from ligand binding.}\label{2ds}
\end{figure}

The free energy landscapes for the four mABP simulations are shown in Figure \ref{2ds}. The tightly coupled CVs project the configuration space onto the diagonal of the CV space. 

\begin{figure}[h]
\includegraphics[width=\columnwidth]{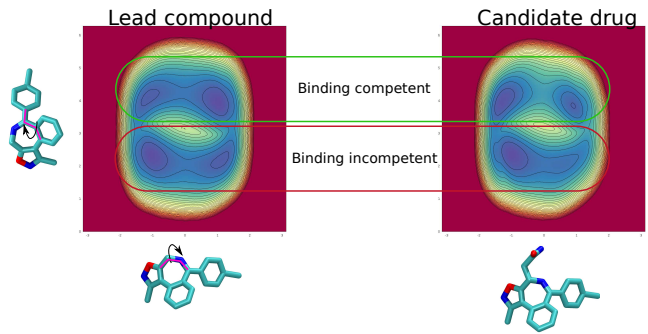}
\caption{Landscape of the azepine pucker. The CV were dihedrals, as indicated pictorially. The crystallographic geometry is given by the upper left minimum on the landscapes. Simulations were 300 ns long and swissparam was used to parameterize the ligands. Contours are every $2k_BT$.}\label{conf}
\end{figure}

In Figure \ref{conf} we show the free energy landscape of the ring puckering for both the lead and final compounds from reference \onlinecite{toy}. Indeed, the immediate and testable hypothesis suggested by these simulations is that this puckering behavior could be controlled to improve potency. The compound on the right side of figure \ref{conf} is CPI-0610, which is the compound that has moved into clinical trials. Of course, binding is not the only consideration when the goal is development of a drug, and this compound was ultimately selected for its favorable pharmacokinetics and toxicological profiles.


\subsection{Definition of the RMSD}
\begin{figure}[h]
\includegraphics[width=\columnwidth]{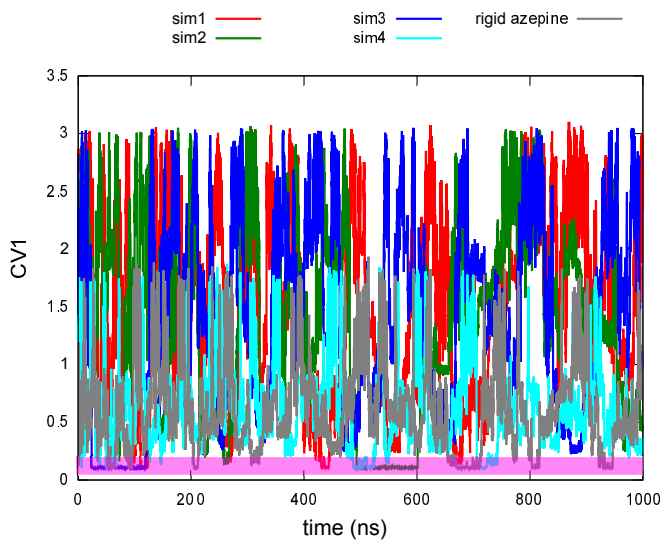}
\caption{Traces of CV1 for all simulations. The pink bar highlights the crystallographic state. See text for details.}\label{trace}
\end{figure}

To avoid division by zero when the RMSD is zero, we write the RMSD as
\[ RMSD(x,x0) = \left( \frac{1}{N}\sum_{i=1}^{N} (x(i)-x_0(i))^2 +0.01\right)^{1/2} \]
which is offset from zero by $0.01$. $x$ are the atomic positions of the atoms in the CV and $x_0$ are the reference positions for the atoms. The lower bound for this RMSD is $0.1$. A trace of CV1 for all trajectories shows that the RMSD is $0.1$ when the crystallographic state is found (Figure \ref{trace}). Sim1 started unbound in spherical restraint, sim2 started bound and used the spherical restraint, sim3 started unbound and used spherical restraint, sim4 started bound and used cylindrical restraint. The rigid azepine simulation started bound and used the cylindrical restraint.

\section{Supplementary Material}
Supplementary material provides details on the hill functions, MPI calls in GROMACS, convergence results for alanine dipeptide, free energy landscapes for ligand binding simulations, free energy of azepine conformations, and the definition of our RMSD coordinate.

\section{Acknowledgements}
All computations were performed on the Van Andel Research Institute compute cluster, except for the timings on EC2. This work was supported in part by the Van Andel Research Institute and research grants from the National Institutes of Health to S.B.R. (CA181343).


\end{document}